\documentclass[12pt,preprint]{aastex}

\slugcomment{To appear in $The$ $Astrophysical$ $Journal$}

\shorttitle{Investigation for the enrichment pattern in HE
0338-3945} \shortauthors{Cui et al.}

\begin{document}
\title {Investigation for the enrichment pattern of the element abundances in r+s star HE 0338-3945: a special r-II star?}
\author{Wen-Yuan Cui\altaffilmark{1}, Jiang Zhang, Zi-Zhong Zhu and Bo Zhang\altaffilmark{2}}

\affil{Department of Physics, Hebei Normal University, 113 Yuhua
Dong Road, Shijiazhuang 050016, P.R.China; \\Hebei Advanced Thin
Films Laboratory, shijiazhuang 050016, P.R.China}

\altaffiltext{1}{E-mail address: cuiwenyuan@hebtu.edu.cn}
\altaffiltext{2}{Corresponding author. E-mail address:
zhangbo@hebtu.edu.cn}

\begin{abstract}
The very metal-poor star HE 0338-3945 shows a double-enhanced
pattern of the neutron-capture elements. The study to this sample
could make people gain a better understanding of s- and r-process
nucleosynthesis at low metallicity. Using a parametric model, we
find that the abundance pattern of the neutron-capture elements
could be best explained by a binary system formed in a molecular
cloud, which had been polluted by r-process material. The observed
abundance pattern of C and N can be explained by an AGB model
\citep{kar07}. Combing with the parameters obtained from
\citet{cui06}, we suggest that the initial mass of the AGB
companion is most likely to be about $2.5M_\odot$, which excludes
the possibility of forming a type-1.5 supernova. By comparing with
the observational abundance pattern of CS 22892-052, we find that
the dominating production of O should accompany with the
production of the heavy r-process elements of r+s stars. Similar
to r-II stars, the heavy r-process elements are not produced in
conjunction with all the light elements from Na to Fe group. The
abundance pattern of the light and r-process elements for HE
0338-3945 is very close to the pattern of the r-II star CS
22892-052. So, we suggest that this star HE 0338-3945 should be a
special r-II star.
\end{abstract}


\section{Introduction}
The two neutron-capture processes, i.e. the (slow) s-process and
the (rapid) r-process, occur under different physical conditions
and are therefore likely to arise in different astrophysical
sites. The dominant site of the s-process is thought to be the
asymptotic giant branch (AGB) phase in low- and intermediate-mass
stars \citep{bus99}. The site or sites of the r-process are not
known, although suggestions include the $\nu$-driven wind of Type
II supernovae \citep{woo92,woo94}, the mergers of neutron stars
\citep{lat74,ros00}, accretion-induced collapse
\citep[AIC;][]{qia03}, and Type 1.5 supernovae
\citep{ibe83,zij04}. The neutron-capture elements are composed of
some pure r-process, some pure s-process, and some mixed-parentage
isotopes. As a result, when the solar system's abundances are
separated into the contributions from the s-process and the
r-process, some elements are mostly contributed by the r-process,
such as Eu, and some by the s-process, such as Ba. Therefore, Eu
is commonly referred to as an ``r-process element'', and Ba as an
``s-process element''.

Observations for metal-poor stars with metallicities lower than
[Fe/H] $=-2.5$ enriched in neutron-capture elements have revealed
the solar r-process pattern, while only a few cases of highly
r-process-enhanced stars \citep[hereafter "r-II"
stars;][]{sne96,sne03,cay01,hil02}, have been noted. Despite their
considerable metal deficiency, these stars seem to have experienced
an r-process that barely differs from the sum of r-processes that
enriched the pre-solar nebula. This has led to suggestions that
r-process production may be independent of the initial metallicity
of the site, especially for the heavier n-capture elements
\citep[$Z\geq 56$;][]{cow95,sne96,sne00,nor97}.

It is puzzling that several stars show enhancements of both
r-process and s-process elements \citep[r+s stars
hereafter;][]{hil00,coh03}, as they require pollution from both an
AGB star and a supernova. The origin of the abundance
peculiarities of the r+s stars is not clear, and many scenarios
have been presented \citep{jon06}. \citet{qia03} proposed a
scenario for the creation of r+s stars. Firstly, some s-process
material is accreted from an AGB star, which turns into a white
dwarf, then, during the evolution of the system, the white dwarf
accretes matter from the polluted star and suffers an AIC to a
neutron star. The $\nu$-driven wind produces an r-process, which
also pollutes the companion. A possible problem, as these authors
mentioned, is the still uncertain nucleosynthesis in
accretion-induced collapse, which may or may not produce the
r-process. Another possible r+s scenario is that the AGB star
transfers s-rich matter to the observed star but does not suffer a
large mass loss, and at the end of the AGB phase the degenerate
core of the low-metallicity, high-mass AGB star may reach the
Chandrasekhar mass, leading to a Type 1.5 supernova \citep{zij04}.
Such suggestion can explain both the enhancement pattern and the
metallicity dependence of the double-enhanced halo stars. There is
another scenario of the origin for the double-enhanced halo
stars. In this picture, the formation of a binary system of
low-mass stars was triggered by a supernova that polluted and
clumped a nearby molecular cloud. Subsequently, the observed star,
which is already strongly enhanced in r-process elements, receives large
amounts of s-process elements from the initially more massive star
that underwent the AGB phase and turns into the double-enhanced
star \citep{aoki02,del04,bar05,gal05,iva05}.

The nucleosynthesis of neutron-capture elements for CEMP
(carbon-enhanced metal-poor) stars can be investigated by the
abundance pattern of r+s stars. Recently, an analysis of the element
abundances for the CEMP star HE 0338-3945 \citep{jon06} showed
that it is rich in both s- and r-elements. \citet{jon06} reported
that this object locates near to the main sequence
turnoff with metallicity of
[Fe/H] $=-2.42$. They concluded that the observed heavy element
abundances of HE 0338-3945 could not be well fit by a scaled solar
r-process pattern nor by a scaled solar s-process pattern. It is a
challenging problem to quantitatively understand of the origins of neutron-capture
elements in the double-enhanced halo stars. Although some of the basic tools for this task
were presented several years ago, the origins of the neutron-capture
elements in the double-enhanced halo stars, especially r-process
elements, are not clear, and the characteristics of the s-process
nucleosynthesis in the AGB stars are not ascertained. Clearly, the
study of element abundances in these objects is important for
investigation of the origin of neutron-capture elements in these
objects and in our Galaxy. One might hope that a clarification of
the origin of r+s stars may shed some light on the general
questions concerning the sites of r- and s-processes.

It is interesting to adopt the parametric model for metal-poor stars
presented by \citet{aoki01} and developed by \citet{zha06} to study
the physical conditions that could reproduce the observed abundance
pattern found in such type stars. In this paper, we investigate the
characteristics of the nucleosynthesis pathway that produces the
special abundance ratios of the r- and s-rich object HE 0338-3945
using the AGB parametric model. The calculated results are presented
in Sect.\ 2, where we also discuss the characteristics of the
s-process nucleosynthesis and the possible origin of their r-process
elements. Conclusions are given in Sect.\ 3.

\section{Results and Discussion}
By comparing the observed abundances pattern with the predicted s- and
r-process contributions, we explore the origin of the heavy elements
in HE 0338-3945. We adopt the parametric model for
metal-poor stars presented by \citet{zha06}. The abundance of the $i$th element
 in the envelope of a star can be calculated as follows:
\begin{equation}
N_{i}(Z)=C_{s}N_{i,\ s}+C_rN_{i,\ r}10^{[Fe/H]} ,
\end{equation}
where $Z$ is the metallicity of the star, $N_{i,\ s}$ is the
abundance of the $i$th element produced by the s-process in the AGB
star and $N_{i,\ r}$ is the abundance of the $i$th element produced
by the r-process (per Si $=10^6$ at $Z=Z_\odot$), whereas, $C_s$ and $C_r$
are the component coefficients that correspond to contributions from
the s-process and the r-process, respectively. The extremely
metal-poor star CS 22892-052 merits special attention, because this
star has an extremely large overabundance of neutron-capture elements
relative to iron with [Fe/H] $=-3.1$. Many
authors \citep{cow99,sne96,sne00,sne03,nor97,pfe97} suggested
that the abundance patterns of the heavier ($Z\geq 56$) stable
neutron-capture elements in very metal-poor stars are consistent
with the solar system r-process abundance distribution, and this
concordance breaks down for the lighter neutron-capture elements in
the range of $40<Z <56$ \citep{sne00}. \citet{zha02} reported
that when the abundances of the lighter elements in CS 22892-052 are
multiplied by a factor of $1/0.4266$, the abundance distributions
obtained for both heavier and lighter neutron-capture elements are
in accordance with the solar system r-process pattern. This star
could have abundances that well reflect the nucleosynthesis of a
single supernova \citep{fie02}, so the adopted abundances of nuclei
$N_{i,\ r}$ in equation $(1)$ are taken as the solar system
r-process abundances \citep{arl99} for the elements heavier than Ba, and
for the lighter nuclei we use solar system r-process
abundances multiplied by a factor of 0.4266. There are four
parameters in our model for r+s stars, such as
the neutron exposure per thermal pulse $\Delta\tau$, the overlap
factor $r$, the component coefficient of the s-process $C_s$, and
the component coefficient of the r-process $C_r$. Using the observed
data in the sample star HE 0338-3945 \citep{jon06}, the parameters
in the model can be obtained from the parametric approach.

Our best-fit results are shown in
figure 1. For HE 0338-3945, the curves produced by our
model are consistent with the observed abundances for almost all the
21 heavy elements within the error limits. The good agreement between
the model results and the observations provides strong support for
the validity and reliability of the derived neutron-capture nucleosynthesis
parameters. The overlap factor $r$
is a fundamental parameter in the AGB model. At solar metallicity,
\citet{gal98} found that the value of the overlap factor $r$ is
between $0.4$ and $0.7$ in their standard evolution model for
low-mass ($1.5-3.0M_\odot$) AGB stars. The overlap factor deduced
for HE 0338-3945 is $r = 0.40$, which lies in the above
range. The overlap factor for different initial-mass AGB
stars as a function of metallicity has been presented by
\citet{cui06} (see their Figure 1). Considering its metallicity,
we suggest that the mass of the primary star (former AGB star)
is less than $3.0M_\odot$, i.e. it lies between $2.0$ and
$3.0M_\odot$.

The neutron exposure per pulse $\Delta\tau$ is another fundamental
parameter. \citet{zha06} have deduced neutron
exposures per pulse for other s-enhanced metal-poor stars, and found they lie
between $0.45$ and $0.88$ mbarn$^{-1}$. The neutron exposure
deduced for HE 0338-3945 is $\Delta\tau=0.77$ mbarn$^{-1}$,
which also lies in above range. In the case of multiple subsequent
exposures, the mean neutron exposure is given by
$\tau_0=-\Delta\tau/$ln$r$. We note that the
value of $\tau_{0}=2.92(\frac{T_{9}}{0.348})^{1/2}$ mbarn$^{-1}$
($T_{9}=0.1$, in units of 10$^{9}$ K) for HE 0338-3945 is
significantly greater than that of
$\tau_{0}=(0.30\pm0.01)(\frac{T_{9}}{0.348})^{1/2}$ mbarn$^{-1}$,
which is best fit the solar system abundances \citep{kap89}. In fact,
the higher mean neutron exposure favors the more amount of
much heavier elements produced, such as Ba and Pb
etc. For the s-only star CS 30322-023, \citet{cui07} obtained its
mean neutron exposure, i.e. $\tau_{0}=2.38(\frac{T_{9}}{0.348})^{1/2}$
mbarn$^{-1}$, which is lower than that of HE
0338-3945. Based on the observation results \citep{jon06,mas06}, we
can find that the abundance ratios of [hs/ls] $=1.36$ (where hs
denotes the `heavy' s-process elements, such as Ba, La, Ce and ls
denotes the `light' s-process elements, such as Sr, Y, Zr) in HE
0338-3945 is larger than that of [hs/ls] $=0.79$ in CS 30322-023.
This is consistent with the relation between the two mean neutron
exposure for these two stars. [Pb/hs] is particularly useful in investigating the
efficiency of the s-process site \citep{str05}.
Using both the convective model and the radiative model,
\citet{cui06} have calculated [Pb/hs] versus [Fe/H]
for different initial stellar mass. Considering the [Pb/hs]$=0.82$ and [Fe/H] $=-2.42$
for HE 0338-3945, we suggest that the mass of the former AGB
star is about $2.5M_\odot$.

For CEMP-s stars the most likely scenario is the pollution by
mass transfer from a more massive AGB companion in a binary
system, the later has undergone AGB nucleosynthesis and become a white dwarf.
The s-process elements and C abundances in HE
0338-3945 are a result of pollution from the dredged-up material
in the former AGB star. The measured [s/Fe] refers to the average
s-process material in the AGB star after dilution by mixing with
the envelope of the companion that is now the
extrinsic star. Other parameters deduced for HE 0338-3945 are
$C_s=0.0050$ and $C_r=61.0$. During the AGB evolution, the
convective He shell and the envelope of the star are
overabundant in heavy elements by factors $f_{shell}$ and
$f_{env,\ 1}$, respectively. The approximate relation between
$f_{env,\ 1}$ and $f_{shell}$ is
\begin{equation}
f_{env,\ 1}\approx\frac{\Delta M_{dr}}{M^e_1}f_{shell} ,
\end{equation}
where $\Delta$$M_{dr}$ is the total mass dredged up from the He
shell into the envelope of the AGB star and $M^{e}_{1}$ is the
envelope mass of the AGB star. For a given s-process element, the
overabundance factor $f_{env,\ 2}$ in the extrinsic star's envelope can
be approximately related to the overabundance factor $f_{env,\ 1}$
by
\begin{equation}
 f_{env,\ 2}\approx\frac{\Delta M_{2}}{M^e_2}f_{env,\ 1}
 \approx\frac{\Delta M_2}{M^e_2}\frac{\Delta M_{dr}}{M^e_1}f_{shell} ,
\end{equation}
where $\Delta$$M_{2}$ is the amount of matter accreted by the extrinsic star,
$M^{e}_{2}$ is the envelope mass of the star. The
component coefficient $C_{s}$ is computed from the relation
\begin{equation}
 C_s =\frac{f_{env,\ 2}}{f_{shell}}\approx\frac{\Delta M_2}{M^e_2}\frac{\Delta
 M_{dr}}{M^e_1} .
\end{equation}

It is found that substantial enhancements of carbon and nitrogen with respect to iron are common
among CEMP stars \citep{coh03,iva05,jon06}. AGB nucleosynthesis
models for low initial mass ($< 2.5M_\odot$) show that carbon is produced by the $3\alpha$ reaction
during thermal pulses, but
not nitrogen because it is burned during the same helium
shell flashes, while for higher mass AGB stars the dredged-up carbon  effectively converts
into nitrogen by CN-cycling at the
bottom of the convective envelope \citep[hot bottom burning,
HBB;][]{pol08}. Detailed evolution models of AGB stars \citep{kar07}
showed that HBB sets in at significantly lower mass at low
metallicity ($3.0M_\odot$ at [Fe/H] $= -2.3$) than that at solar
metallicity (around $5.0M_\odot$). In figure 2, we show the
enhancements of N and C relative to iron in HE 0338-3945
and compare them with the predicted abundance ratios for AGB
models according to \citet{kar07} with $Z=0.0001$. The
possible abundance ratios of the companion polluted by a former AGB
star with different initial mass are plotted as solid lines. From this
figure we can see that AGB stars with masses between $2.0$ and
$3.0M_\odot$ produce nitrogen and carbon in the right amounts to
account for the observed abundances, after accretion of the material
by a low-mass companion and subsequent dilution in its envelope.
For $2.5M_\odot$, the factor $\frac{\Delta
M_{dr}}{M^e_1}$ is about $1/6$ \citep{kar07} and comparing
the observation abundance of C and N with the calculated results of
\citet{kar07}, the dilution factor
$\frac{\Delta M_2}{M^e_2}$ for HE 0338-3945 is about $1/33$ (see figure 2).
The value of
$C_s=0.0051$ is deduced from equation $(4)$, which is close to our
calculated value, i.e. $C_s=0.0050$, for HE 0338-3945 using the
parametric model. As the observational abundance ratios of [C/Fe]
$=2.13\pm0.15$ and [N/Fe] $=1.55\pm0.17$ \citep{jon06}, we believe
that HBB does not happen in the AGB companion, i.e. the
initial mass of the AGB companion is less than $3.0M_\odot$
\citep{kar07} with its metallicity considered, which is in agreement with
the obtained results. It should be mentioned that the C and N
abundances come from the \citet{jon06} 1D LTE analysis based on
the CH and CN molecule lines, however, which are uncertain, due
to the strong temperature sensitivity of the CH and CN molecule
formation. Preliminary results for the 3D corrections may amount to approximately
$-0.5$ to $-0.3$ and $-0.5$ dex for carbon and nitrogen, respectively \citep{asp01,asp04,jon06}.
The results, when the 3D corrections of $-0.5$ dex for carbon and nitrogen are considered,
are also shown in figure 2. We can see form this figure that the
3D effects do not influence our results.

\citet{zij04} suggested that the strong metallicity dependence of
mass loss during the AGB phase leads to a steeper initial-final mass
relation for low-metalliciy stars, that is, for a given initial
mass, the final mass is higher for metal-poor stars. Therefore, the
core of, e.g., a metal-poor star with [Fe/H] $=-3.0$ having an
initial mass of $4.0M_\odot$ would reach the Chandrasekhar mass,
leading to a ``Type 1.5" supernova. In such a supernova, r-process
nucleosynthesis might occur, and the surface of the
companion star observed today could have been polluted. Our results imply clearly
that HE 0338-3945 should be polluted by a
$2.5M_\odot$ AGB star, which could not cause a ``Type 1.5" supernova. The
component coefficient of the r-process calculated for HE 0338-3945
is $C_r=61.0$, which means that this star and its AGB companion
should form in a molecular cloud that had already  been polluted by r-process
material.

Apart from carbon and nitrogen, we concentrate on oxygen
enrichments as possible tracers of the origin of CEMP stars.
\citet{jon06} derived a super-solar oxygen abundance of [O/Fe] $=
+1.40$ in the halo star HE 0338-3945. In figure 3, we show the
enhancements of O and C relative to iron observed in HE
0338-3945 and compare these to the abundance ratios in the
material lost by AGB stars with [Fe/H] $= -2.3$ according to the
\citet{kar07}. The abundance ratios of the companion polluted by a former AGB
star with different initial mass are plotted as solid lines.
This figure shows that AGB stars with masses
between $1.25$ and $6.0M_\odot$ do not produce enough
O to account for the observed abundance. This means that O doesn't
mainly come from the former AGB star, it must have another
origin. It is noteworthy that the O abundance derived by \citet{jon06}
based on 1D LTE analysis of the O I triple lines, which are known
to be sensitive to the NLTE effects. \citet{jon06} estimated the
NLTE effects were to be $\sim-0.1$ dex. However, recently,
\citet{fab09} have redone the O I NLTE calculations with new
electron collisional data, which leads to much larger NLTE
effects, about $-0.3$ to $-0.6$ dex for the given stellar
parameters. However, the exact corrections depend on the still
uncertain hydrogen collisional cross-sections. For this star, we
estimate that the NLTE correction of O is of the order of $-0.5$
dex, and the result is also shown in figure 3. From this figure,
we can see that the O abundance in HE 0338-3945 is still not
reproduced by the AGB model, even if the NLTE corrections have
been considered.

The observed results for HE 0338-3945 \citep{jon06} are shown as
filled circles in figure 4, where the red solid line represents the
results calculated by the parametric model, while the black solid
line represents the results given by \citet{kar07} for $2.5M_\odot$
AGB model adopting a factor $\frac{\Delta M_2}{M^e_2}$ of 1/33. It
can be seen that there is good agreement between the observation
data on C, N and these of the AGB model within the error limits. It
is interesting to simultaneously analyse the observed light and
heavy elements to study the physical conditions that could reproduce
the observed abundance pattern found in this star. Indeed, some very
metal-poor stars, such as CS 22892-052, have high abundance ratios
of heavier neutron-capture elements, e.g. Eu \citep{sne03}. CS
22892-052 possibly record the abundance patterns produced by the
r-process and is called as an r-II star. In this star, the
abundances of neutron-capture elements are in remarkable agreement
with the scaled solar system r-process pattern \citep{sne03}. This
is the first determination of the overall abundance pattern that
could represent the yields of the r-process. The abundances in this
star must have been printed and enhanced by the r-process elements
and other elements which are produced in conjunction with the
r-process. In order to investigate the light elements pattern in
this r+s star, we use CS 22892-052 as an r-process only star, and
plot its observed element abundances normalized to Eu in this
figure, too \citep[see the blue dash dot line;][]{sne03}. It is
believed that Eu is mainly made by the r-process. For the two stars,
HE 0338-3945 and CS 22892-052, shown in this figure, have almost
identical abundance pattern from N to Cu including O, but the C and
heavy s-process elements, such as Ba and Pb, differ greater than
$0.8$ dex, which indicates that the origin of C and s-process
elements is AGB stars. Because the overabundance of O and C for HE
0338-3945, could not be explained simultaneously by any AGB star
(see figure 3), we can conclude that the production of O should
accompany with the production of the heavy r-process elements for
this r+s star. We note that though the overabundance of O for HE
0338-3945 is larger than that for CS 22892-052, but when considering
its also larger overabundance of Eu, the abundance pattern of O and
r-process elements for HE 0338-3945 is still close to that for the
r-II star CS 22892-052. It can be found that the heavy r-process
elements are not produced in conjunction with all the elements from
Na to Fe group elements, which is similar to r-II stars, too
\citep{qia07}. Combined the calculated results of neutron-capture
elements, we find that the abundance pattern of the r-process and
light elements in HE 0338-3945 is vary close to that of the r-II
star CS 22892-052. This implies that this star should be a special
r-II star? In this figure, we do not attempt to correct the C, N and
O abundance in HE 0338-3945 for 3D or NLTE effects quantitatively
because of the still uncertain H collisions. Although the
corrections are expected to decrease their abundances, our
conclusions are not significantly influenced.

It is important to discuss the uncertainty of the derived parameters
using the method presented by \citet{aoki01}. Figures 5 and 6 show
the calculated ratios log$\varepsilon$(Sr), log$\varepsilon$(Ba),
log$\varepsilon$(Pb) and log$\varepsilon$(Eu) as a function of the
neutron exposure $\Delta\tau$ in a model with $r=0.40$ and as a
function of overlap factor $r$ with a fixed neutron exposure
$\Delta\tau=0.77$ mbarn$^{-1}$. These are compared with the observed
abundance ratios in HE 0338-3945. There is only a region of overlap
in figure 5, $\Delta\tau=0.77^{+0.01}_{-0.04}$ mbarn$^{-1}$, in
which all the observed ratios referred above can be accounted for.
The bottom panel in figure 5 displays the reduced $\chi^2$ value
calculated in our model, and there is a minimum $\chi^2 =1.008$ at
$\Delta\tau=0.77$ mbarn$^{-1}$; the neutron exposure is constrained
quite well. The abundance ratios log$\varepsilon$(Sr),
log$\varepsilon$(Ba), log$\varepsilon$(Pb) and log$\varepsilon$(Eu)
are insensitive to the overlap factor $r$ and allow for a wider
range, i.e. $0.3 < r < 0.42$.

\section{ Conclusions}
The chemical abundances of the extremely r- and s-rich star HE
0338-3945 are an excellent test bed to set new constraints on models
of neutron-capture process nucleosynthesis at low metallicity. The
neutron-capture element abundance pattern of HE 0338-3945 could be
explained by a binary system formed in a molecular cloud that had
ever been polluted by r-process material. Based on the good fit
using the AGB models \citep{kar07} and parametric model
\citep{zha06} to match the observed abundance pattern including the
abundances of C and N, we deduce that the progenitor mass of the AGB
star is about $2.5M_\odot$. The more massive companion underwent an
AGB phase, and produced lots of C, N and s-process isotopes.
Subsequently, the envelope of HE 0338-3945 is enriched with the
dredged-up material from the former AGB star. The initial mass of
the AGB companion is less than $4.0M_\odot$, which excludes the
possibility of forming a Type 1.5 supernova. Overall, the predicted
abundances for the r/s-process fit well the observed abundance
patterns of HE 338-3945, for the first (Sr and Zr) and second peaks
(Ba and La), as well as the third peak of the s-process (Pb). We
find that the production of O should accompany with the production
of the heavy r-process elements for r+s stars and the contribution
of O in HE 0338-3945 from the AGB stars is swamped by the
pre-enrichment of O. Similar to r-II stars, the heavy r-process
elements are not produced in conjunction with all the elements from
Na to Fe group elements. The abundance pattern of the r-process and
light elements in HE 0338-3945 is very close to that of the r-II
star CS 22892-052. This implies that this star may be a special r-II
star.

Clearly, it is important for future studies to determine the
relation between r-II and r+s stars, whether the r-pattern in r-II
stars can explain the abundance distribution of the r-material in
r+s stars. More in-depth theoretical and observational studies of
r-rich and r+s stars will reveal the characteristics of the r- and
s-process at low metallicity and the history of the enrichment of
neutron-capture elements in the early Galaxy.

\acknowledgments

The authors thank the anonymous referees for insightful comments,
containing very relevant scientific advice which improved this paper
greatly. We also thank Dr. Jian-Rong Shi for the careful review of
the manuscript and for fruitful discussion. This work has been
supported by the National Natural Science Foundation of China under
grant no.10847119, 10673002, 10973006 and 10778616, by the Natural
Science Foundation of Hebei Proince under Grant no. A2009000251, by
the Science Foundation of Hebei Provincial Education Department
under Grant no. 2008127, by the Science Foundation of Hebei Normal
University under Grant no. L2007B07.

\clearpage

\begin{figure}
\epsscale{1.10} \plotone{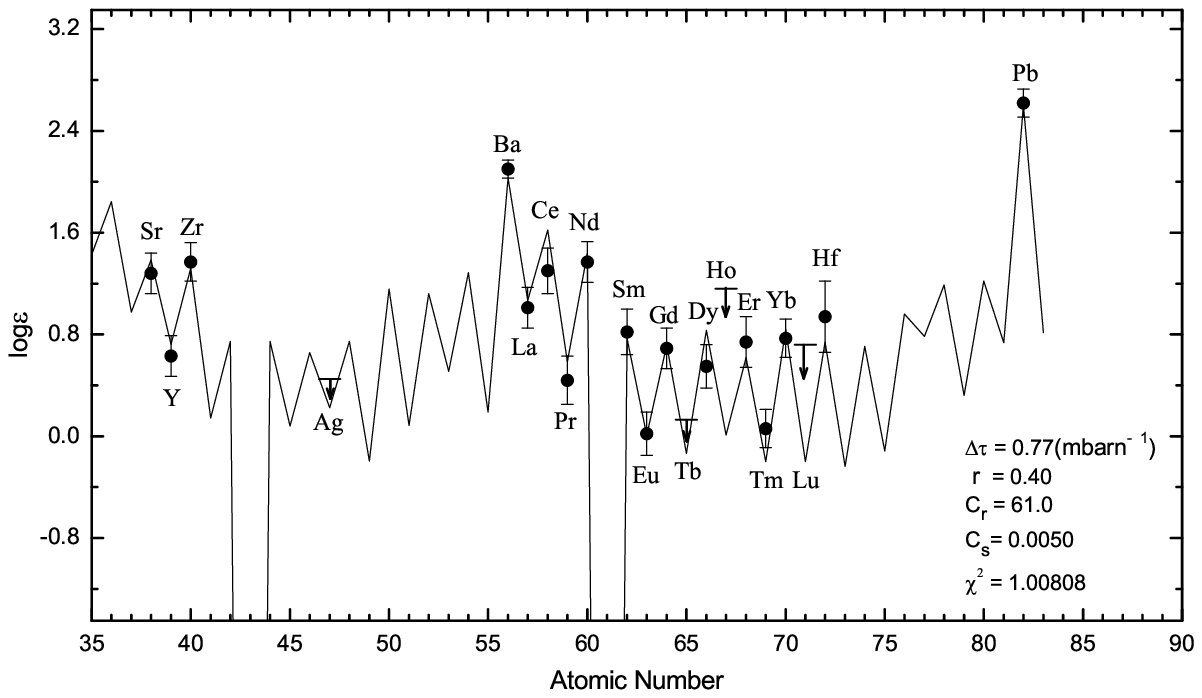} \caption{{Best fit to
observational results of metal-deficient star HE 0338-3945. The
filled circles with appropriate error bars and downward arrows
denote the observed element abundances, the solid lines represent
predictions from s-process calculations, in which the r-process
contribution is considered simultaneously. The standard unit of
$\Delta\tau$ is mbarn$^{-1}$. }\label{fig1}}
\end{figure}

\begin{figure}
\epsscale{1.15} \plotone{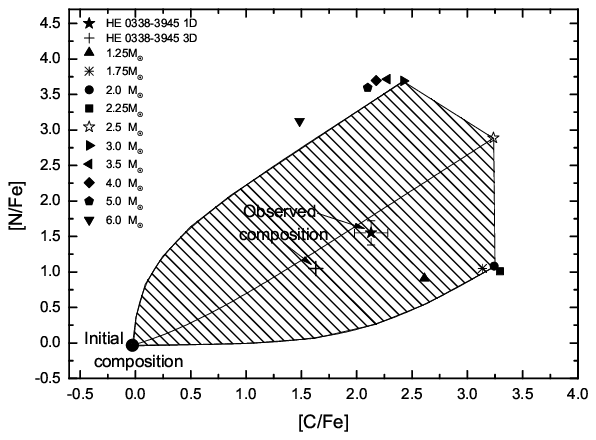} \caption{{Abundance ratios of N
and C with respect to Fe as observed in HE 0338-3945 with
appropriate error bars \citep[filled asterisk, taken from][]{jon06},
and predicted by \citet{kar07} AGB stars at [Fe/H] $= -2.3$, with
labels indicating the initial masses. The cross represents the
abundance ratios which have been corrected by 3D effects. The
initial composition is scaled from solar. The hatched ellipsoid is
the region of the plot where the N and C abundances from the AGB
companion should lie to ensure that the observed abundances are
reproduced after dilution. The unfilled asterisk represents possible
former AGB star with initial mass $2.5M_\odot$ computed by
\citet{kar07} models.}\label{fig2}}
\end{figure}

\begin{figure}
\epsscale{1.15} \plotone{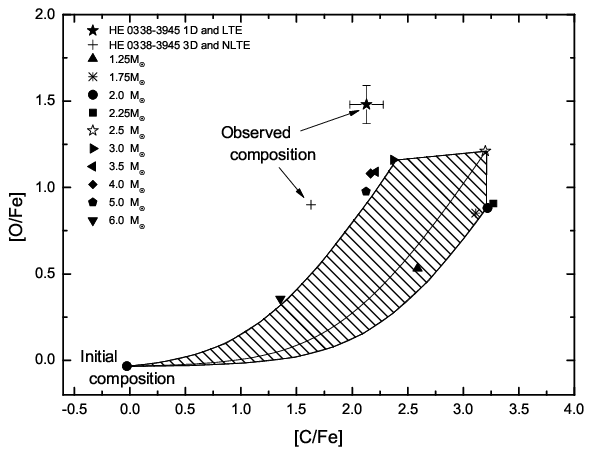} \caption{{Abundance ratios of O
and C with respect to Fe as observed in HE 0338-3945 with
appropriate error bars \citep[filled asterisk, taken from][]{jon06},
and predicted by \citet{kar07} AGB stars at [Fe/H] $= -2.3$, with
labels indicating the initial masses. The cross represents the
abundance ratios which have considered the NLTE effects for O and 3D
effects for C, respectively. The initial composition is scaled from
solar. The hatched ellipsoid is the region of the plot where the
observed abundances of O and C can be reproduced by the abundances
after dilution from the AGB companion predicted by \citet{kar07}
models.}\label{fig3}}
\end{figure}

\begin{figure}
\epsscale{1.10} \plotone{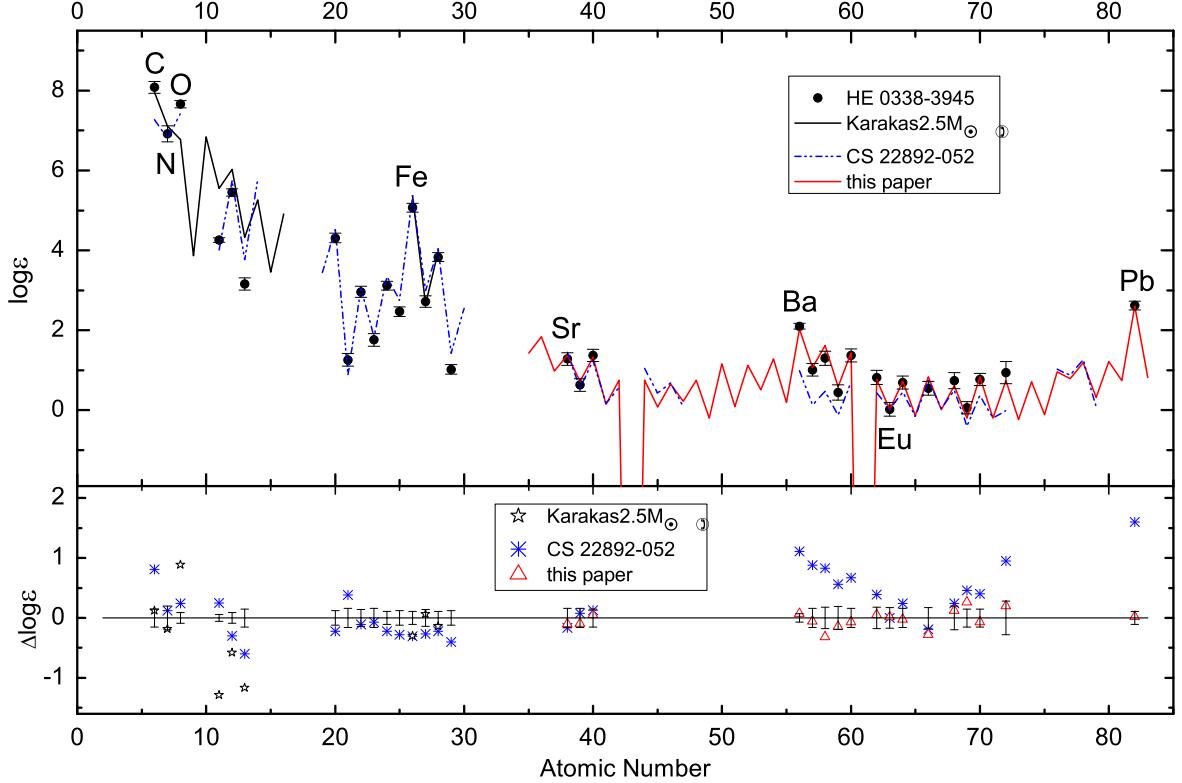} \caption{{The abundance pattern
of HE 0338-3945 (filled circles and downward arrows) compare with
predictions from s-process calculations of parametric AGB model with
r-process contribution considered (red solid line), predictions from
\citet{kar07} $2.5M_\odot$ star model with [Fe/H] $=-2.3$ (black
solid line), and the abundance pattern of a r-II star CS 22892-052
\citep[adopted from][]{sne03} scaled to the Eu abundance of HE
0338-3945 (blue dash dot line). The estimated absolute error bars
are shown. The bottom panel displays the difference, defined as
$\Delta$log$\varepsilon$(X) $\equiv$
log$\varepsilon$(X)$_{obs}-$log$\varepsilon$(X)$_{calc}$, and upper
limits are not shown. Note that none of the displayed C, N and O
abundances have been corrected for 3D or NLTE effects.}\label{fig4}}
\end{figure}

\begin{figure}
\epsscale{1.00} \plotone{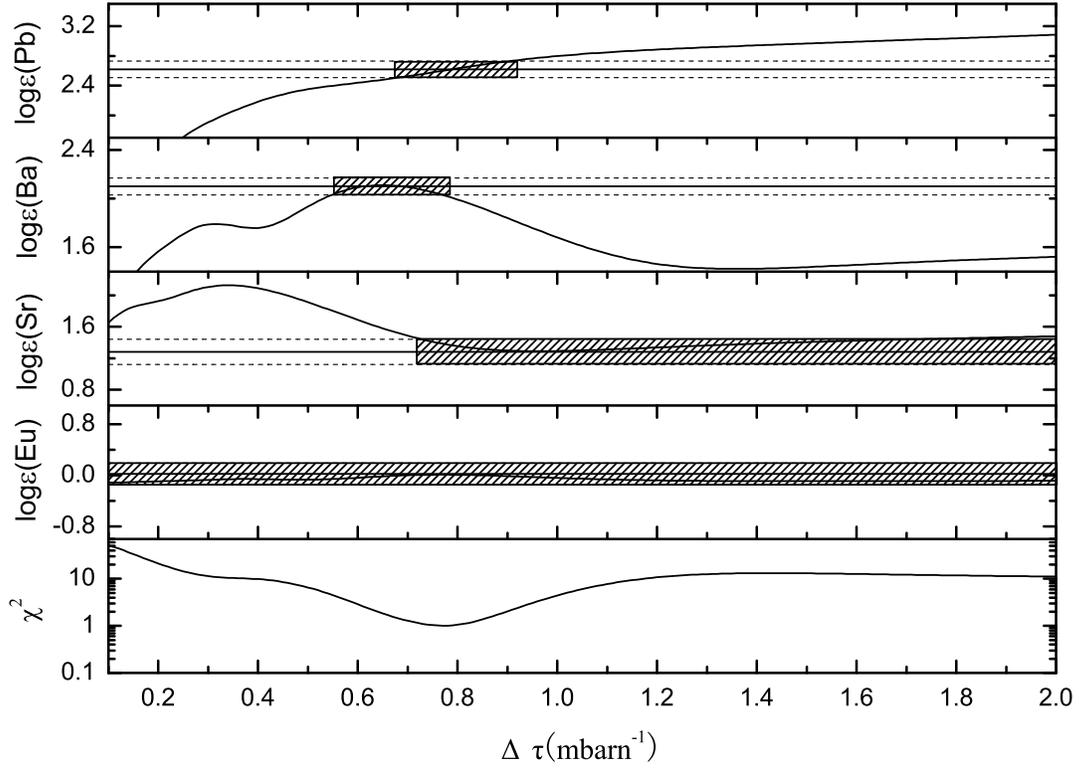} \caption{{Logarithm abundances
log$\varepsilon$(Pb), log$\varepsilon$(Ba), log$\varepsilon$(Sr) and
log$\varepsilon$(Eu) in r+s star HE 0338-3945, and reduced $\chi^2$
(bottom) as a function of the neutron exposure $\Delta\tau$ computed
by a model with $C_r=61.0$, $C_s=0.0050$ and $r=0.40$. These are
compared with the observed abundances of HE 0338-3945 adopted from
\citet{jon06}. }\label{fig5}}
\end{figure}

\begin{figure}
\epsscale{1.00} \plotone{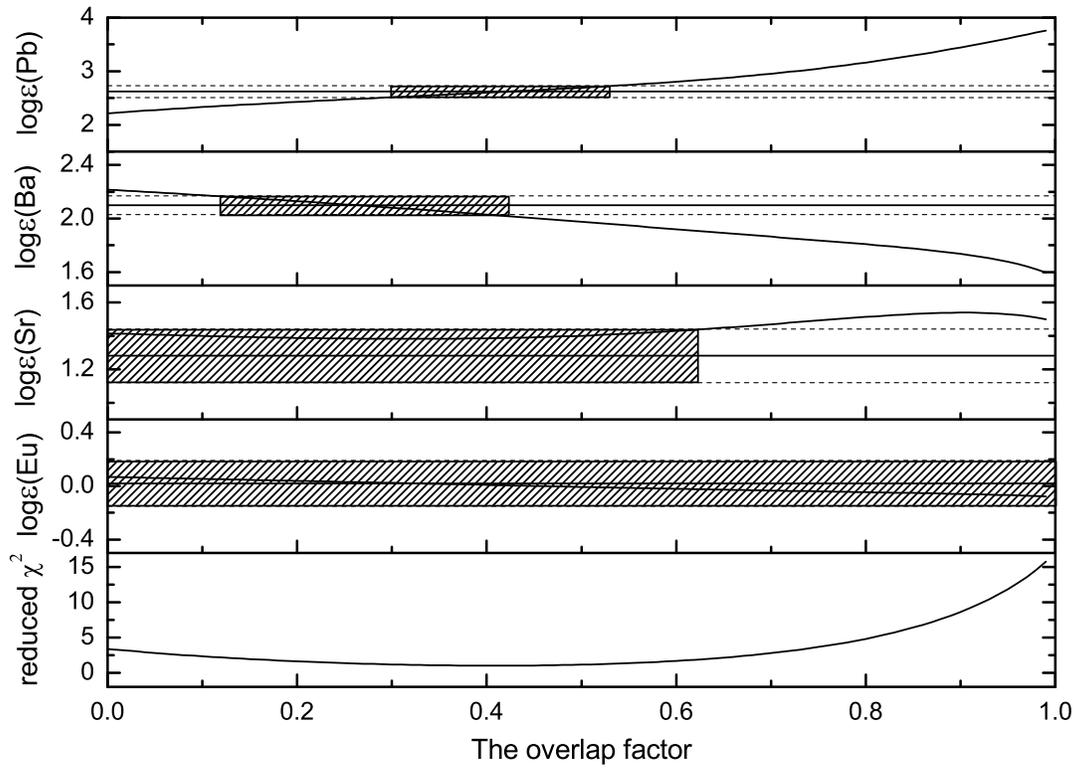} \caption{{Same as those in Fig. 5
but as a function of the overlap factor $r$ in a model with
$\Delta\tau=0.77$ mbarn$^{-1}$. }\label{fig6}}
\end{figure}


\end{document}